\documentclass[a4paper]{report}
\usepackage[utf8]{inputenc}
\usepackage[T1]{fontenc}
\usepackage{RJournal}
\usepackage{amsmath,amssymb,array}
\usepackage{booktabs}

\newcommand{\rev}[1]{\textcolor{black}{#1}}
\makeatletter
\renewcommand{\thesection}{\@arabic\c@section}
\makeatother

\begin{document}

\begin{article}

\title{Robust and Efficient Optimization Using a Marquardt-Levenberg Algorithm
with R Package marqLevAlg}

\author{
by Viviane Philipps, Boris P. Hejblum, Mélanie Prague, Daniel Commenges and Cécile Proust-Lima
}

\maketitle

\abstract{
Implementations in R of classical general-purpose algorithms for local optimization generally have two major limitations which cause difficulties in applications to complex problems: too loose convergence criteria and too long calculation time. By relying on a Marquardt-Levenberg algorithm (MLA), a Newton-like method particularly robust for solving local optimization problems, we provide with marqLevAlg package an efficient and general-purpose local optimizer which (i) prevents convergence to saddle points by using a stringent convergence criterion based on the relative distance to minimum/maximum in addition to the stability of the parameters and of the objective function; and (ii) reduces the computation time in complex settings by allowing parallel calculations at each iteration. We demonstrate through a variety of cases from the literature that our implementation reliably and consistently reaches the optimum (even when other optimizers fail), and also largely reduces computational time in complex settings through the example of maximum likelihood estimation of different sophisticated statistical models.
}

\hypertarget{introduction}{%
\section{Introduction}\label{introduction}}

Optimization is an essential task in many computational problems. In
statistical modelling for instance, in the absence of analytical
solution, maximum likelihood estimators are often retrieved using 
iterative optimization algorithms which locally solve the problem from given starting values.

Steepest descent algorithms are among the most famous general \rev{local}
optimization algorithms. They generally consist in updating parameters
according to the steepest gradient (gradient descent) possibly scaled by
the Hessian in the Newton (Newton-Raphson) algorithm or an approximation
of the Hessian based on the gradients in the quasi-Newton algorithms
(e.g., Broyden-Fletcher-Goldfarb-Shanno - BFGS). Newton-like
algorithms have been shown to provide good convergence properties
\citep{joe_numerical_2003} and were demonstrated in particular to behave
better than Expectation-Maximization (EM) algorithms in several contexts
of Maximum Likelihood Estimation, such as the random-effect models
\citep{lindstrom_newtonraphson_1988} or the latent class models
\citep{proust_estimation_2005}. Among Newton methods, the
Marquardt-Levenberg algorithm, initially proposed by Levenberg
\citep{levenberg_method_1944} then Marquardt
\citep{marquardt_algorithm_1963}, combines BFGS and gradient descent
methods to provide a more robust optimization algorithm. \rev{As other Newton methods, Marquardt-Levenberg algorithm is designed to find a local optimum of the objective function from given initial values. When dealing with multimodal objective functions, it can thus converge to local optimum, and needs to be combined with a grid search to retrieve the global optimum. }

The R software includes multiple solutions for \rev{local} and \rev{global} optimization
tasks (see CRAN task View on \ctv{Optimization} \citep{optimization}).
In particular the \code{optim}
function in \CRANpkg{base} R offers different algorithms for
general purpose optimization, and so does \CRANpkg{optimx}, a more
recent package extending \code{optim} \citep{nash_2011}. Numerous
additional packages are available for different contexts, from nonlinear
least square problems (including some exploiting Marquardt-Levenberg
idea like \CRANpkg{minpack.lm} \citep{elzhov_2016}
\rev{and \CRANpkg{nlmrt} \citep{nlmrt_2016}}) to stochastic
optimization and algorithms based on the simplex approach. However,
R software could benefit from a general-purpose R implementation of Marquardt-Levenberg algorithm.

We present here an R implementation of the Marquardt-Levenberg algorithm in the package \CRANpkg{marqLevAlg} which
relies on a stringent convergence criterion based on the first and second derivatives to avoid loosely convergence
\citep{prague:hal-00717566} and includes (from version 2.0.1) parallel
computations within each iteration to speed up
convergence. This implementation is particularly dedicated to complex
settings, that is when a large number of parameters are optimized, and/or the computation of the objective function is time-consuming. The parallel computations speed up the procedure, and the stringent convergence criterion prevents false convergences on the flat regions of the objective function obtained with convergence criteria based on the function stability.

Section 2 and 3 describe the algorithm and the implementation,
respectively. Then Section 4 provides an example of call with the
estimation of a linear mixed model. A benchmark of the package is
reported in Section 5 with the performances of parallel implementation.
\rev{Performances of} Marquardt-Levenberg algorithm implementation \rev{are also challenged in Section 6 using a variety of simple and complex examples from the literature, and compared with other optimizers}. Finally Section 7 concludes.

\hypertarget{methodology}{%
\section{Methodology}\label{methodology}}

\hypertarget{the-marquardt-levenberg-algorithm}{%
\subsection{The Marquardt-Levenberg
algorithm}\label{algorithm}}

The Marquardt-Levenberg algorithm (MLA) can be used for any problem
where a function \rev{\(\mathcal{F}(\theta)\)} has to be minimized (or
equivalently, function \(\mathcal{L}(\theta)\)= -
\(\mathcal{F}(\theta)\) has to be maximized) according to a set of \(m\)
unconstrained parameters \(\theta\), as long as the second derivatives
of \(\mathcal{F}(\theta)\) exist. In statistical applications for
instance, the objective function is the deviance to be minimized or the
log-likelihood to be maximized.

Our improved MLA iteratively updates the vector \(\theta^{(k)}\) from a
starting point \(\theta^{(0)}\) until convergence using the following
formula at iteration \(k+1\):

\[\theta^{(k+1)}=\theta^{(k)}-\delta_{k} \left(\tilde{H}(\mathcal{F}(\theta^{(k)}))\right)^{-1}\nabla(\mathcal{F}(\theta^{(k)}))\]

where \(\theta^{(k)}\) is the set of parameters at iteration \(k\),
\(\nabla(\mathcal{F}(\theta^{(k)}))\) is the gradient of the objective
function at iteration \(k\), and \(\tilde{H}(\mathcal{F}(\theta^{(k)}))\) is the Hessian
matrix \(H(\mathcal{F}(\theta^{(k)}))\) where the diagonal terms are replaced by
\(\tilde{H}(\mathcal{F}(\theta^{(k)}))_{ii}=H(\mathcal{F}(\theta^{(k)}))_{ii}+\lambda_k[(1-\eta_k)|H(\mathcal{F}(\theta^{(k)}))_{ii}|+\eta_k \text{tr}(H(\mathcal{F}(\theta^{(k)})))]\).
In the original MLA the Hessian matrix is inflated by a scaled identity
matrix. Following \citet{fletcher_modified_1971} we consider a refined
inflation based on the curvature. The diagonal inflation of our improved
MLA makes it an intermediate between the steepest descent method and the
Newton method. The parameters \(\delta_k\), \(\lambda_k\) and \(\eta_k\)
are scalars specifically determined at each iteration \(k\). Parameter
\(\delta_k\) is fixed to 1 unless the objective function is not reduced,
in which case a line search determines the locally optimal step length.
Parameters \(\lambda_k\) and \(\eta_k\) are internally modified in order
to ensure that (i) \(\tilde{H}(\mathcal{F}(\theta^{(k)}))\) be definite-positive at each
iteration \(k\), and (ii) \(\tilde{H}(\mathcal{F}(\theta^{(k)}))\) approaches \(H(\mathcal{F}(\theta^{(k)}))\)
when \(\theta^{(k)}\) approaches \(\hat{\theta}\).

When the problem encounters a unique solution, the minimum is reached
whatever the chosen initial values.

\hypertarget{sec:criteria}{%
\subsection{Stringent convergence criteria}\label{sec:criteria}}

As in any iterative algorithm, convergence of MLA is achieved when
convergence criteria are fullfilled. In \CRANpkg{marqLevAlg} package,
convergence is defined according to three criteria:

\begin{itemize}
\item
  parameters stability:
  \(\sum_{j=1}^{m} \left(\theta_{j}^{(k+1)}-\theta_{j}^{(k)}\right)^2 < \epsilon_a\)
\item
  objective function stability:
  \(|\mathcal{F}^{(k+1)} - \mathcal{F}^{(k)}| < \epsilon_b\)
\item
  relative distance to minimum/maximum (RDM):
  \(\frac{\nabla(\mathcal{F}(\theta^{(k)})) \left(H(\mathcal{F}(\theta^{(k)}))\right)^{-1} \nabla(\mathcal{F}(\theta^{(k)})) }{m} < \epsilon_d\)
\end{itemize}

\rev{The original Marquardt-Levenberg algorithm \citep{marquardt_algorithm_1963} and its implementations \citep{elzhov_2016,nlmrt_2016} consider the two first criteria (as well as a third one based on the angle between the objective function and its gradient). Yet these criteria, which are also used in many other iterative algorithms, do not ensure a convergence towards an actual optimum. They only ensure the convergence towards a saddle point. We thus chose to complement the parameter and objective function stability by the relative distance to minimum/maximum. As it requires the Hessian matrix to be invertible, it prevents from any convergence to a saddle point, and is thus essential to ensure that an optimum is truly reached.}
When the Hessian is not invertible, RDM is set to 1+\(\epsilon_d\) and
convergence criteria cannot be fullfilled.

Although it constitutes a relevant convergence criterion in any
optimization context, RDM was initially designed for log-likelihood
maximization problems, that is cases where \(\mathcal{F}(\theta)\)= -
\(\mathcal{L}(\theta)\) with \(\mathcal{L}\) the log-likelihood. In that
context, RDM can be interpreted as the ratio between the numerical error
and the statistical error
\citep[\citet{prague2013nimrod}]{commenges_rvs_2006}.

The three thresholds \(\epsilon_a\), \(\epsilon_b\) and \(\epsilon_d\)
can be adjusted, but values around \(0.0001\) are usually sufficient to
guarantee a correct convergence. In some complex log-likelihood
maximisation problems for instance, \citet{prague2013nimrod} showed that
the RDM convergence properties remain acceptable providing
\(\epsilon_d\) is below 0.1 (although the lower the better).

\hypertarget{derivatives-calculation}{%
\subsection{Derivatives calculation}\label{derivatives}}

MLA update relies on first (\(\nabla(\mathcal{F}(\theta^{(k)}))\)) and
second (\(H(\mathcal{F}(\theta^{(k)}))\)) derivatives of the objective function
\(\mathcal{F}(\theta^{(k)})\) at each iteration k. The gradient and the
Hessian may sometimes be calculated analytically but in a general
framework, numerical approximation can become necessary. In
\CRANpkg{marqLevAlg} package, in the absence of analytical gradient
computation, the first derivatives are computed by central finite
differences. In the absence of analytical Hessian, the second
derivatives are computed using forward finite differences. The step of
finite difference for each derivative depends on the value of the
involved parameter. It is set to \(\max(10^{-7},10^{-4}|\theta_j|)\) for
parameter \(j\).

When both the gradient and the Hessian are to be numerically computed,
numerous evaluations of \(\mathcal{F}\) are required at each iteration:

\begin{itemize}
\item
  \(2\times m\) evaluations of \(\mathcal{F}\) for the numerical
  approximation of the gradient function;
\item
  \(\dfrac{m \times (m+1)}{2}\) evaluations of \(\mathcal{F}\) for the
  numerical approximation of the Hessian matrix.
\end{itemize}

The number of derivatives thus grows quadratically with the number \(m\)
of parameters and calculations are per se independent as done for
different vectors of parameters \(\theta\).

When the gradient is analytically calculated, only the second
derivatives have to be approximated, requiring \(2 \times m\)
independent calls to the gradient function. In that case, the complexity
thus linearly increases with \(m\).

In both cases, and especially when each calculation of derivative is
long and/or \(m\) is large, parallel computations of independent
\(\mathcal{F}\) evaluations becomes particularly relevant to speed up
the estimation process.

\hypertarget{sec:loglik}{%
\subsection{Special case of a log-likelihood
maximization}\label{sec:loglik}}

When the optimization problem is the maximization of the log-likelihood
\(\mathcal{L}(\theta)\) of a statistical model according to parameters
\(\theta\), the Hessian matrix of the
\(\mathcal{F}(\theta) = - \mathcal{L}(\theta)\) calculated at the
optimum \(\hat{\theta}\),
\(\mathcal{H}(\mathcal{F}({\hat{\theta}})) = - \dfrac{\partial^2 \mathcal{L}(\theta)}{\partial \theta^2} \bigg|_{\theta = \hat{\theta}}\),
provides an estimator of the Fisher Information matrix. The inverse of
\(\mathcal{H}(\mathcal{F}({\hat{\theta}}))\) computed in the package thus provides an
estimator of the variance-covariance matrix of the optimized vector of
parameters \(\hat{\theta}\).

\hypertarget{implementation}{%
\section{Implementation}\label{implementation}}

\hypertarget{marqlevalg-function}{%
\subsection{marqLevAlg function}\label{marqlevalgFunction}}

The call of the \code{marqLevAlg} function, or its shorcut \code{mla},
is the following :

\begin{verbatim}
marqLevAlg(b, m = FALSE, fn, gr = NULL, hess = NULL, maxiter = 500,
  epsa = 0.0001, epsb = 0.0001, epsd = 0.0001, digits = 8,
  print.info = FALSE, blinding = TRUE, multipleTry = 25, nproc = 1,
  clustertype = NULL, file = "", .packages = NULL, minimize = TRUE, ...)
\end{verbatim}

Argument \code{b} is the set of initial parameters; alternatively its
length \code{m} can be entered. \code{fn} is the function to optimize;
it should take the parameter vector as first argument, and additional
arguments are passed in \dots . Optional \code{gr} and \code{hess} refer
to the functions implementing the analytical calculations of the
gradient and the Hessian matrix, respectively. \code{maxiter} is the
maximum number of iterations. Arguments \code{epsa}, \code{epsb} and
\code{epsd} are the thresholds for the three convergence criteria
defined in Section \ref{sec:criteria}. \code{print.info} specifies if
details on each iteration should be printed; such information can be
reported in a file if argument \code{file} is specified, and
\code{digits} indicates the number of decimals in the eventually
reported information during optimization. \code{blinding} is an option
allowing the algorithm to go on even when the \code{fn} function returns
NA, which is then replaced by the arbitrary value of \(500,000\) (for
minimization) and -\(500,000\) (for maximization). Similarly, if an
infinite value is found for the chosen initial values, the
\code{multipleTry} option will internally reshape \code{b} (up to
\code{multipleTry} times) until a finite value is got, and the algorithm
can be correctly initialized. The parallel framework is first stated by
the \code{nproc} argument which gives the number of cores and by the
\code{clustertype} argument (see the next section). In the case where
the \code{fn} function depends on R packages, these should be
given as a character vector in the \code{.packages} argument. Finally,
the \code{minimize} argument offers the possibility to minimize or
maximize the objective function \code{fn}; a maximization problem is
implemented as the minimization of the opposite function (\code{-fn}).

\hypertarget{implementation-of-parallel-computations}{%
\subsection{Implementation of parallel
computations}\label{implementationParallel}}

In the absence of analytical gradient calculation, derivatives are
computed in the \code{deriva} subfunction with two loops, one for the first
derivatives and one for the second derivatives. Both loops are
parallelized. The parallelized loops are at most over \(m*(m+1)/2\)
elements for \(m\) parameters to estimate which suggests that the
performance could theoretically be improved with up to \(m*(m+1)/2\)
cores.

When the gradient is calculated analytically, the \code{deriva} subfunction
is replaced by the \code{deriva\_grad} subfunction. It is parallelized in the
same way but the parallelization being executed over \(m\) elements, the
performance should be bounded at \(m\) cores.

In all cases, the parallelization is achieved using the \CRANpkg{doParallel}
and \CRANpkg{foreach} packages. The snow and multicore options of the
\code{doParallel} backend are kept, making the parallel option of
\CRANpkg{marqLevAlg} package available on all systems. The user specifies
the type of parallel environment among FORK, SOCK or MPI in argument
\code{clustertype} and the number of cores in \code{nproc}. For
instance, \code{clustertype = "FORK", nproc = 6} will use FORK
technology and 6 cores.

\hypertarget{example}{%
\section{Example}\label{example}}

We illustrate how to use \code{marqLevAlg} function with the maximum
likelihood estimation in a linear mixed model
\citep{laird_random-effects_1982}. Function \code{loglikLMM} available
in the package implements the log-likelihood of a linear mixed model for
a dependent outcome vector ordered by subject (argument \(Y\)) explained
according to a matrix of covariates (argument \(X\)) entered in the same
order as \(Y\) with a Gaussian individual-specific random intercept and
Gaussian independent errors:

\begin{example} 
loglikLMM(b, Y, X, ni) 
\end{example}

Argument \(b\) specifies the vector of parameters with first the
regression parameters (length given by the number of columns in \(X\))
and then the standard deviations of the random intercept and of the
independent error. Finally argument \(ni\) specifies the number of
repeated measures for each subject.

We consider the dataset \code{dataEx} (available in the package) in
which variable \(Y\) is repeatedly observed at time \(t\) for 500
subjects along with a binary variable \(X1\) and a continuous variable
\(X3\). For the illustration, we specify a linear trajectory over time
adjusted for \(X1\), \(X3\) and the interaction between \(X1\) and time
\(t\). The vector of parameters to estimate corresponds to the
intercept, 4 regression parameters and the 2 standard deviations.

We first define the quantities to include as argument in
\code{loglikLMM} function:

\begin{example}
> Y <- dataEx$Y
> X <- as.matrix(cbind(1, dataEx[, c("t", "X1", "X3")], 
+                      dataEx$t * dataEx$X1))
> ni <- as.numeric(table(dataEx$i))
\end{example}

The vector of initial parameters to specify in \code{marqLevAlg} call is
created with the trivial values of 0 for the fixed effects and 1 for the
variance components.

\begin{example}
> binit <- c(0, 0, 0, 0, 0, 1, 1)
\end{example}

The maximum likelihood estimation of the linear mixed model in
sequential mode is then run using a simple call to \code{marqLevAlg}
function for a maximization (with argument \code{minimize = FALSE}):

\begin{example}
> estim <- marqLevAlg(b = binit, fn = loglikLMM, minimize = FALSE, 
+                     X = X, Y = Y, ni = ni)
> estim
 
                   Robust marqLevAlg algorithm                    
 
marqLevAlg(b = binit, fn = loglikLMM, minimize = FALSE, X = X, 
    Y = Y, ni = ni)
 
Iteration process: 
      Number of parameters: 7  
      Number of iterations: 18 
      Optimized objective function: -6836.754  
      Convergence criteria satisfied 
 
Convergence criteria: parameters stability= 3.2e-07 
                    : objective function stability= 4.35e-06 
                    : Matrix inversion for RDM successful 
                    : relative distance to maximum(RDM)= 0 
 
Final parameter values: 
 50.115  0.106  2.437  2.949 -0.376 -5.618  3.015 
 
\end{example}

The printed output \code{estim} shows that the algorithm converged in 18
iterations with convergence criteria of 3.2e-07, 4.35e-06 and 0 for
parameters stability, objective function stability and RDM,
respectively. The output also displays the list of coefficient values at
the optimum. All this information can also be recovered in the
\code{estim} object, where item \code{b} contains the estimated
coefficients.

As mentioned in Section \ref{sec:loglik}, in log-likelihood maximization
problems, the inverse of the Hessian given by the program provides an
estimate of the variance-covariance matrix of the coefficients at the
optimum. The upper triangular matrix of the inverse Hessian is thus
systematically computed in object \code{v}. When appropriate, the
\code{summary} function can output this information with option
\code{loglik = TRUE}. With this option, the summary also includes the
square root of these variances (i.e., the standards errors), the
corresponding Wald statistic, the associated \(p\) value and the 95\%
confidence interval boundaries for each parameter:

\begin{example}
> summary(estim, loglik = TRUE)
 
                   Robust marqLevAlg algorithm                    
 
marqLevAlg(b = binit, fn = loglikLMM, minimize = FALSE, X = X, 
    Y = Y, ni = ni)
 
Iteration process: 
      Number of parameters: 7  
      Number of iterations: 18 
      Optimized objective function: -6836.754  
      Convergence criteria satisfied 
 
Convergence criteria: parameters stability= 3.2e-07 
                    : objective function stability= 4.35e-06 
                    : Matrix inversion for RDM successful 
                    : relative distance to maximum(RDM)= 0 
 
Final parameter values: 
   coef SE.coef        Wald P.value   binf   bsup
 50.115   0.426 13839.36027   0e+00 49.280 50.950
  0.106   0.026    16.02319   6e-05  0.054  0.157
  2.437   0.550    19.64792   1e-05  1.360  3.515
  2.949   0.032  8416.33202   0e+00  2.886  3.012
 -0.376   0.037   104.82702   0e+00 -0.449 -0.304
 -5.618   0.189   883.19775   0e+00 -5.989 -5.248
  3.015   0.049  3860.64370   0e+00  2.919  3.110
 
\end{example}

The exact same model can also be estimated in parallel mode using FORK
implementation of parallelism (here with two cores):

\begin{example}
> estim2 <- marqLevAlg(b = binit, fn = loglikLMM, minimize = FALSE, 
+                      nproc = 2, clustertype = "FORK", 
+                      X = X, Y = Y, ni = ni)
\end{example}

It can also be estimated by using analytical gradients (provided in
gradient function \code{gradLMM} with the same arguments as
\code{loglikLMM}):

\begin{example}
> estim3 <- marqLevAlg(b = binit, fn = loglikLMM, gr = gradLMM, 
+                      minimize = FALSE, X = X, Y = Y, ni = ni)
\end{example}

In all three situations, the program converges to the same maximum as
shown in Table \ref{tab:fit} for the estimation process and in Table
\ref{tab:estim} for the parameter estimates. The iteration process is
identical when using the either the sequential or the parallel code
(number of iterations, final convergence criteria, etc). It necessarily
differs slightly when using the analytical gradient, as the computations
steps are not identical (e.g., here it converges in 15 iterations rather
than 18) but all the final results are identical.

\begin{table}[ht]
\centering
\begin{tabular}{lrrr}
  \toprule
 & Object estim & Object estim2 & Object estim3 \\ 
  \midrule
Number of cores & 1 & 2 & 1 \\ 
  Analytical gradient & no & no & yes \\ 
  Objective Function & -6836.754 & -6836.754 & -6836.754 \\ 
  Number of iterations & 18 & 18 & 15 \\ 
  Parameter Stability & 3.174428e-07 & 3.174428e-07 & 6.633702e-09 \\ 
  Likelihood stability & 4.352822e-06 & 4.352822e-06 & 9.159612e-08 \\ 
  RDM & 1.651774e-12 & 1.651774e-12 & 2.935418e-17 \\ 
   \bottomrule
\end{tabular}
\caption{Summary of the estimation process of a linear mixed model using marqLevAlg function run either in sequential mode with numerical gradient calculation (object estim), parallel mode with numerical gradient calculation (object estim2), or sequential mode with analytical gradient calculation (object estim3).} 
\label{tab:fit}
\end{table}

\begin{table}[ht]
\centering
\begin{tabular}{rrrrrrr}
  \toprule & \multicolumn{2}{r}{Object estim} & \multicolumn{2}{r}{Object estim2} & \multicolumn{2}{r}{Object estim3} \\ \midrule
   & Coef & SE &  Coef & SE &  Coef & SE \\ \midrule
Parameter 1 & 50.1153 & 0.4260 & 50.1153 & 0.4260 & 50.1153 & 0.4260 \\ 
  Parameter 2 & 0.1055 & 0.0264 & 0.1055 & 0.0264 & 0.1055 & 0.0264 \\ 
  Parameter 3 & 2.4372 & 0.5498 & 2.4372 & 0.5498 & 2.4372 & 0.5498 \\ 
  Parameter 4 & 2.9489 & 0.0321 & 2.9489 & 0.0321 & 2.9489 & 0.0321 \\ 
  Parameter 5 & -0.3764 & 0.0368 & -0.3764 & 0.0368 & -0.3764 & 0.0368 \\ 
  Parameter 6 & -5.6183 & 0.1891 & -5.6183 & 0.1891 & 5.6183 & 0.1891 \\ 
  Parameter 7 & 3.0145 & 0.0485 & 3.0145 & 0.0485 & 3.0145 & 0.0485 \\ 
   \bottomrule
\end{tabular}
\caption{Estimates (Coef) and standard error (SE) of the parameters of a linear mixed model fitted using marqLevAlg function run either in sequential mode with numerical gradient calculation (object estim), parallel mode with numerical gradient calculation (object estim2), or sequential mode with analytical gradient calculation (object estim3).} 
\label{tab:estim}
\end{table}

\section{Benchmark}\label{benchmark}

We aimed at evaluating and comparing the performances of the
parallelization in some time consuming examples. We focused on three
examples of sophisticated models from the mixed models area estimated by
maximum likelihood. These examples rely on packages using three
different languages, thus illustrating the behavior of \CRANpkg{marqLevAlg}
package with a program exclusively written in R (\CRANpkg{JM},
\citet{rizopoulos_jm_2010}), and programs including Rcpp
(\pkg{CInLPN}, \citet{tadde_dynmod}) and Fortran90
(\CRANpkg{lcmm}, \citet{proust-lima_lcmm_2017}) languages widely used in
complex situations.

We first describe the generated dataset on which the benchmark has been
realized. We then intoduce each statistical model and associated
program. Finally, we detail the results obtained with the three
programs. Each time, the model has been estimated sequentially and with
a varying number of cores in order to provide the program speed-up. We
used a Linux cluster with 32 cores machines and 100 replicates to assess
the variability. Codes and dataset used in this section are available at
\url{https://github.com/VivianePhilipps/marqLevAlgPaper.}

\hypertarget{simulated-dataset}{%
\subsection{Simulated dataset}\label{simulatedDataset}}

We generated a dataset of \(20,000\) subjects having repeated
measurements of a marker \code{Ycens} (measured at times \code{t}) up to
a right-censored time of event \code{tsurv} with indicator that the
event occured \code{event}. The data were generated according to a 4
latent class joint model \citep{proust-lima_joint_2014}. This model
assumes that the population is divided in 4 latent classes, each class
having a specific trajectory of the marker defined according to a linear
mixed model with specific parameters, and a specific risk of event
defined according to a parametric proportional hazard model with
specific parameters too. The class-specific linear mixed model included
a basis of natural cubic splines with 3 equidistant knots taken at times
5, 10 and 15, associated with fixed and correlated random-effects. The
proportional hazard model included a class-specific Weibull risk
adjusted on 3 covariates: one binary (Bernoulli with 50\% probability)
and two continous variables (standard Gaussian, and Gaussian with mean
45 and standard deviation 8). The proportion of individuals in each
class is about 22\%, 17\%, 34\% and 27\% in the sample.

Below are given the five first rows of the three first subjects:

\begin{example}
   i class X1        X2       X3 t    Ycens     tsurv event
1  1     2  0 0.6472205 43.42920 0 61.10632 20.000000     0
2  1     2  0 0.6472205 43.42920 1 60.76988 20.000000     0
3  1     2  0 0.6472205 43.42920 2 58.72617 20.000000     0
4  1     2  0 0.6472205 43.42920 3 56.76015 20.000000     0
5  1     2  0 0.6472205 43.42920 4 54.04558 20.000000     0
22 2     1  0 0.3954846 43.46060 0 37.95302  3.763148     1
23 2     1  0 0.3954846 43.46060 1 34.48660  3.763148     1
24 2     1  0 0.3954846 43.46060 2 31.39679  3.763148     1
25 2     1  0 0.3954846 43.46060 3 27.81427  3.763148     1
26 2     1  0 0.3954846 43.46060 4       NA  3.763148     1
43 3     3  0 1.0660837 42.08057 0 51.60877 15.396958     1
44 3     3  0 1.0660837 42.08057 1 53.80671 15.396958     1
45 3     3  0 1.0660837 42.08057 2 51.11840 15.396958     1
46 3     3  0 1.0660837 42.08057 3 50.64331 15.396958     1
47 3     3  0 1.0660837 42.08057 4 50.87873 15.396958     1
\end{example}

\hypertarget{statistical-models}{%
\subsection{Statistical models}\label{statisticalModels}}

\hypertarget{joint-shared-random-effect-model-for-a-longitudinal-marker-and-a-time-to-event-package-jm}{%
\subsubsection{Joint shared random effect model for a longitudinal
marker and a time to event: package
JM}\label{jm}}

The maximum likelihood estimation of joint shared random effect models
has been made available in R with the \CRANpkg{JM} package
\citep{rizopoulos_jm_2010}. The implemented optimization functions are
\code{optim} and \code{nlminb}. We added the \code{marqLevALg} function
for the purpose of this example. We considered a subsample of the
simulated dataset, consisting in \(5,000\) randomly selected subjects.

The joint shared random effect model is divided into two submodels
jointly estimated:

\begin{itemize}
\item
  a linear mixed submodel for the repeated marker \(Y\) measured at
  different times \(t_{ij}\) (\(j=1,...,n_i\)): \[\begin{split}
  Y_{i}(t_{ij})  &= \tilde{Y}_{i}(t_{ij}) + \varepsilon_{ij}\\
 &= X_i(t_{ij}) \beta + Z_i(t_{ij}) u_i + \varepsilon_{ij}
  \end{split}\]
\end{itemize}

where, in our example, \(X_i(t)\) contained the intercept, the class
indicator, the 3 simulated covariates, a basis of natural cubic splines
on time \(t\) (with 2 internal knots at times 5 and 15) and the
interactions between the splines and the time-invariant covariates,
resulting in 20 fixed effects. \(Z_i(t)\) contained the intercept and
the same basis of natural cubic splines on time \(t\), and was
associated with \(u_i\), the 4-vector of correlated Gaussian random
effects. \(\varepsilon_{ij}\) was the independent Gaussian error.

\begin{itemize}
\item
  a survival submodel for the right censored time-to-event: \[
   \alpha_i(t) = \alpha_0(t) \exp(X_{si}\gamma + \eta \tilde{Y}_{i}(t))
  \]
\end{itemize}

where, in our example, the vector \(X_{si}\), containing the 3 simulated
covariates, was associated with the vector of parameters \(\gamma\); the
current underlying level of the marker \(\tilde{Y}_{i}(t)\) was
associated with parameter \(\eta\) and the baseline hazard
\(\alpha_{0}(t)\) was defined using a basis of B-splines with 1 interior
knot.

The length of the total vector of parameters \(\theta\) to estimate was
40 (20 fixed effects and 11 variance component parameters in the
longitudinal submodel, and 9 parameters in the survival submodel).

One particularity of this model is that the log-likelihood does not have
a closed form. It involves an integral over the random effects (here, of
dimension 4) which is numerically computed using an adaptive
Gauss-Hermite quadrature with 3 integration points for this example.

As package \CRANpkg{JM} includes an analytical computation of the gradient,
we ran two estimations: one with the analytical gradient and one with
the numerical approximation to compare the speed up and execution times.

\hypertarget{latent-class-linear-mixed-model-package-lcmm}{%
\subsubsection{Latent class linear mixed model: package
lcmm}\label{lcmm}}

The second example is a latent class linear mixed model, as implemented
in the \code{hlme} function of the \CRANpkg{lcmm} R package. The
function uses a previous implementation of the Marquardt algorithm coded
in Fortran90 and in sequential mode. For the purpose of this
example, we extracted the log-likelihood computation programmed in
Fortran90 to be used with \CRANpkg{marqLevAlg} package.

The latent class linear mixed model consists in two submodels estimated
jointly:

\begin{itemize}
\item
  a multinomial logistic regression for the latent class membership
  (\(c_i\)):
\end{itemize}

\[\mathbb{P}(c_i = g) = \frac{\exp(W_{i} \zeta_g)}{\sum_{l=1}^G \exp(W_{i} \zeta_l)}  ~~~~~~~~~~~~~ \text{with      }  g=1,...,G \]
where \(\zeta_G=0\) for identifiability and \(W_{i}\) contained an
intercept and the 3 covariates.

\begin{itemize}
\item
  a linear mixed model specific to each latent class \(g\) for the
  repeated outcome \(Y\) measured at times \(t_{ij}\) (\(j=1,...,n_i\)):
\end{itemize}

\[ Y_i(t_{ij} | c_i = g) =  X_i(t_{ij}) \beta_g + Z_i(t_{ij}) u_{ig} + \varepsilon_{ij}\]

where, in this example, \(X_i(t)\) and \(Z_i(t)\) contained an
intercept, time \(t\) and quadratic time. The vector \(u_{ig}\) of
correlated Gaussian random effects had a proportional variance across
latent classes, and \(\varepsilon_{ij}\) were independent Gaussian
errors.

The log-likelihood of this model has a closed form but it involves the
logarithm of a sum over latent classes which can become computationally
demanding. We estimated the model on the total sample of \(20,000\)
subjects with 1, 2, 3 and 4 latent classes which corresponded to 10, 18,
26 and 34 parameters to estimate, respectively.

\hypertarget{multivariate-latent-process-mixed-model-package-cinlpn}{%
\subsubsection{Multivariate latent process mixed model: package
CInLPN}\label{cinlpn}}

The last example is provided by the \pkg{CInLPN} package, which relies
on the Rcpp language. The function fits a multivariate linear
mixed model combined with a system of difference equations in order to
retrieve temporal influences between several repeated markers
\citep{tadde_dynmod}. We used the data example provided in the package
where three continuous markers \code{L\_1}, \code{L\_2}, \code{L\_3} were
repeatedly measured over time. The model related each marker \(k\)
(\(k=1,2,3\)) measured at observation times \(t_{ijk}\) (\(j=1,...,T\))
to its underlying level \(\Lambda_{ik}(t_{ijk})\) as follows:
\[\text{L}_{ik}(t_{ijk}) = \eta_{0k}+ \eta_{1k} \Lambda_{ik}(t_{ijk}) +\epsilon_{ijk}\]
where \(\epsilon_{ijk}\) are independent Gaussian errors and
\((\eta_0,\eta_1)\) parameters to estimate. Simultaneously, the
structural model defines the initial state at time 0
(\(\Lambda_{ik}(0)\)) and the change over time at subsequent times \(t\)
with \(\delta\) is a discretization step:

\[
\begin{split}
 \Lambda_{ik}(0) &= \beta_{0k} + u_{ik}\\
 \frac{\Lambda_{ik}(t+\delta) - \Lambda_{ik}(t)}{\delta} &= \gamma_{0k} + v_{ik}  + \sum_{l=1}^K a_{kl} \Lambda_{il}(t)
\end{split} 
\]

where \(u_{ik}\) and \(v_{ik}\) are Gaussian random effects.

Again, the log-likelihood of this model that depends on 27 parameters
has a closed form but it may involve complex calculations.

\hypertarget{results}{%
\subsection{Results}\label{results}}

\begin{figure}
  \centering
  \includegraphics[width=1\linewidth]{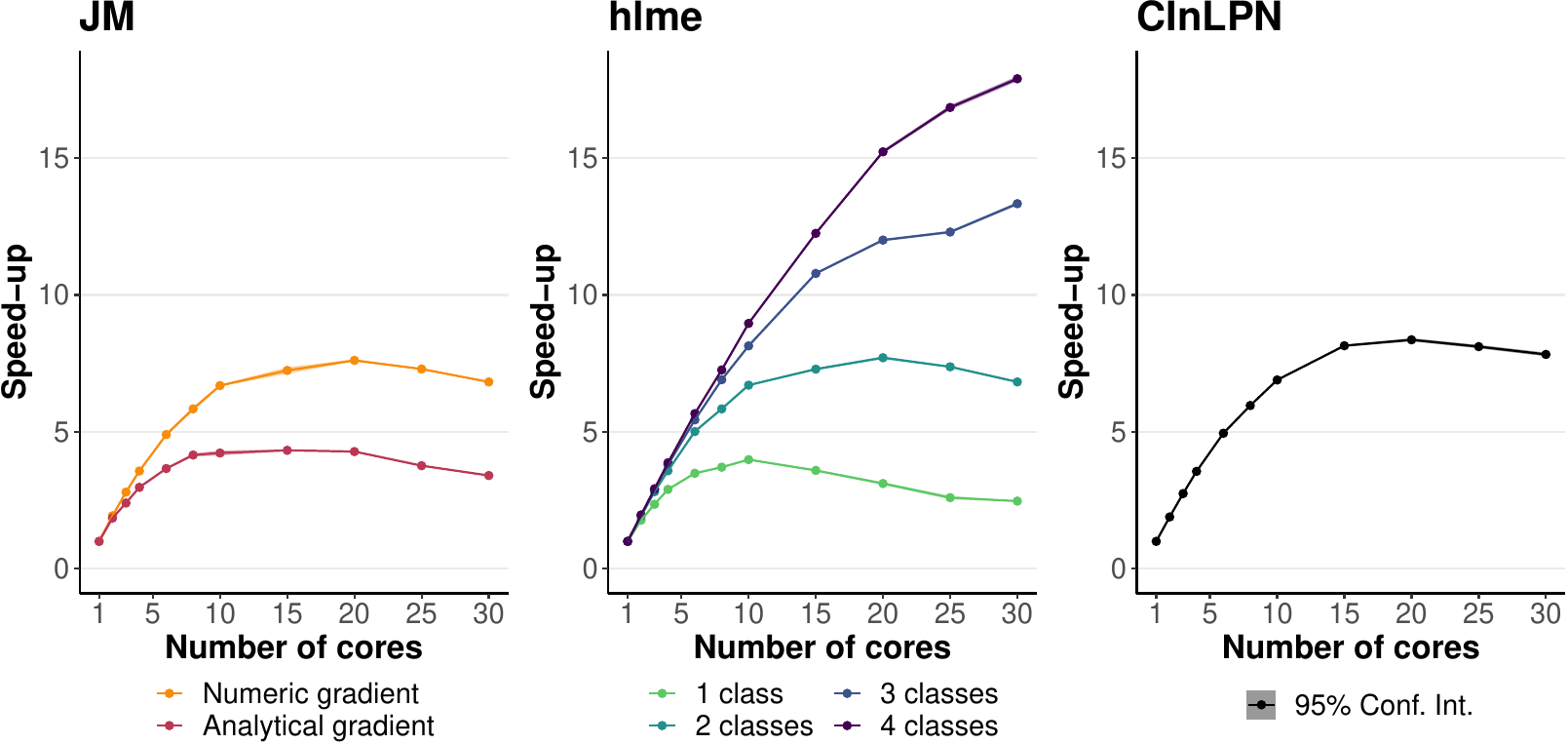} 
  \caption[Speed up performances for the 3 different programs (JM, hlme and CInLPN)]{Speed up performances for the 3 different programs (JM, hlme and CInLPN). Analytic and numeric refer to the analytical and numerical computations of the gradient in JM. The number of parameters was  40 for JM; 10, 18, 26, 34 for hlme with 1, 2, 3, 4 classes, respectively; 27 for CInLPN.}\label{fig:speedup}
\end{figure}

All the models have been estimated with 1, 2, 3, 4, 6, 8, 10, 15, 20, 25
and 30 cores. To fairly compare the execution times, we ensured that
changing the number of cores did not affect the final estimation point
or the number of iterations needed to converge. The mean of the speed up
over the 100 replicates are reported in table \ref{tab:perf} and plotted
in Figure \ref{fig:speedup}.

\begin{table}[ht]
\centering
\begingroup\footnotesize
\begin{tabular}{lrrrrrrr}
  \toprule & \multicolumn{2}{c}{JM} & \multicolumn{4}{c}{hlme} & CInLPN \\ \midrule
   & analytic & numeric & G=1 & G=2 & G=3 & G=4 & \\ \midrule
Number of parameters & 40 & 40 & 10 & 18 & 26 & 34 & 27 \\ 
  Number of iterations & 16 & 16 & 30 & 30 & 30 & 30 & 13 \\ 
  Number of elements in foreach loop & 40 & 860 & 65 & 189 & 377 & 629 & 405 \\ 
  Sequential time (seconds) & 4279 & 14737 & 680 & 3703 & 10402 & 22421 & 272 \\ 
  Speed up with 2 cores & 1.85 & 1.93 & 1.78 & 1.93 & 1.94 & 1.96 & 1.89 \\ 
  Speed up with 3 cores & 2.40 & 2.80 & 2.35 & 2.81 & 2.88 & 2.92 & 2.75 \\ 
  Speed up with 4 cores & 2.97 & 3.57 & 2.90 & 3.58 & 3.80 & 3.87 & 3.56 \\ 
  Speed up with 6 cores & 3.66 & 4.90 & 3.49 & 5.01 & 5.44 & 5.66 & 4.95 \\ 
  Speed up with 8 cores & 4.15 & 5.84 & 3.71 & 5.84 & 6.90 & 7.26 & 5.96 \\ 
  Speed up with 10 cores & 4.23 & 6.69 & 3.98 & 6.70 & 8.14 & 8.96 & 6.89 \\ 
  Speed up with 15 cores & 4.32 & 7.24 & 3.59 & 7.29 & 10.78 & 12.25 & 8.14 \\ 
  Speed up with 20 cores & 4.28 & 7.61 & 3.11 & 7.71 & 12.00 & 15.23 & 8.36 \\ 
  Speed up with 25 cores & 3.76 & 7.29 & 2.60 & 7.37 & 12.30 & 16.84 & 8.11 \\ 
  Speed up with 30 cores & 3.41 & 6.82 & 2.47 & 6.82 & 13.33 & 17.89 & 7.83 \\ 
   \bottomrule
\end{tabular}
\endgroup
\caption{Estimation process characteristics for the 3 different programs (JM, hlme and CInLPN). Analytic and Numeric refer to the analytical and numerical computations of the gradient in JM; G refers to the number of latent classes.} 
\label{tab:perf}
\end{table}

The joint shared random effect model (\code{JM}) converged in 16
iterations after 4279 seconds in sequential mode when using the
analytical gradient. Running the algorithm in parallel on 2 cores made
the execution 1.85 times shorter. Computational time was gradually
reduced with a number of cores between 2 and 10 to reach a maximal speed
up slightly above 4. With 15, 20, 25 or 30 cores, the performances were
no more improved, the speed up showing even a slight reduction, probably
due to the overhead. In contrast, when the program involved numerical
computations of the gradient, the parallelization reduced the
computation time by a factor of almost 8 at maximum. The better speed-up
performances with a numerical gradient calculation were expected since
the parallel loops iterate over more elements.

The second example, the latent class mixed model estimation
(\code{hlme}), showed an improvement of the performances as the
complexity of the models increased. The simple linear mixed model (one
class model), like the joint models with analytical gradient, reached a
maximum speed-up of 4 with 10 cores. The two class mixed model with 18
parameters, showed a maximum speed up of 7.71 with 20 cores. Finally the
3 and 4 class mixed models reached speed-ups of 13.33 and 17.89 with 30
cores and might still be improved with larger resources.

The running time of the third program (CInLPN) was also progressively
reduced with the increasing number of cores reaching the maximal
speed-up of 8.36 for 20 cores.

In these 7 examples, the speed up systematically reached almost 2 with 2
cores, and it remained interesting with 3 or 4 cores although some
variations in the speed-up performances began to be observed according
to the complexity of the objective function computations. This hilights
the benefit of the parallel implementation of MLA even on personal
computers. As the number of cores continued to increase, the speed-up
performances varied a lot. Among our examples, the most promising
situation was the one of the latent class mixed model (with program in
Fortran90) where the speed-up was up to 15 for 20 cores with
the 4 class model.

\hypertarget{comparison-with-other-optimization-algorithms}{%
\section{Comparison with other optimization
algorithms}\label{comparison}}

\hypertarget{other-marquardt-levenberg-implementations}{%
\subsection{Other Marquardt-Levenberg
implementations}\label{otherML}}

The Marquardt-Levenberg algorithm has been previouly implemented in the context of nonlinear least squares problems in \CRANpkg{minpack.lm} and \CRANpkg{nlmrt}. We ran the examples provided in these two packages with \code{marqLevAlg} and compared the algorithms in terms of final solution (that is the residual sum-of-squares) and runtime. Results are shown in supplementary material. Our implementation reached exactly the same value as the two others but performed slower in these simple examples.

We also compared the sensitivity to initial values of \code{marqLevAlg} with \CRANpkg{minpack.lm} using a simple example from \CRANpkg{minpack.lm}. We ran the two implementations of MLA on 100 simulated datasets each one from 100 different starting points (see supplementary material). On the 10000 runs, \code{marqLevAlg} converged in 51.55\% of the cases whereas the \CRANpkg{minpack.lm} converged in 65.98\% of the cases. However, 1660 estimations that converged according to nls.lm criteria were far from the effective optimum. This reduced the proportion of satisfying convergences with \CRANpkg{minpack.lm} to 49.38\% (so similar rate as \code{marqLevAlg}) but more importantly illustrates the convergence to saddle points when using classical convergence criteria. In contrast, all the convergences with \CRANpkg{marqLevAlg} were closed to the effective solution thanks to its stringent RDM convergence criterion.

\hypertarget{examples-from-the-literature}{%
\subsection{Examples from the literature}\label{exliterature}}

We tested our algorithm on 35 optimization problems designed by \citet{more_1981} to test unconstrained optimization software, and compared the \code{marqLevAlg} performances with those of several other optimizers, namely Nelder-Mead, BFGS, conjugate gradients (CG) implemented in the \code{optim} function, L-BFGS-B algorithm from \code{optimParallel}, and \code{nlminb}. Each problem consists of a function to optimize from given starting points. The results are presented in supplementary material in terms of bias between the real solution and the final value of the objective function. Our implementation of MLA converged in almost all the cases (31 out of 35), and provided almost no bias. Except for nlminb which showed similar very good performances, the other algorithms converged at least once very far from the effective objective value. In addition, Nelder-Mead and CG algorithms converged only in approximately half of the cases.

\hypertarget{joint-models}{%
\subsection{Example of complex optimization problem: Maximum Likelihood Estimation of a Joint model for longitudinal and time-to-event data}\label{jm}}

Our implementation is particularly dedicated to complex problems involving many parameters and/or complex objective function calculation. We illustrate here its performances and compare them with other algorithms for the likelihood maximization of a joint model for longitudinal and time-to-event data, as an example of complex objective function optimization.

The \CRANpkg{JM} package (\citet{rizopoulos_jm_2010}), \rev{dedicated to the maximum likelihood estimation of joint models,} includes several optimization algorithms, namely the BFGS of \code{optim} function, and an expectation-maximization (EM) technique internally implemented. It thus offers a nice framework to compare the reliability of MLA to find the
maximum likelihood \rev{in a complex setting} with the reliability of other
optimization algorithms. We used in this comparison the \code{prothro}
dataset described in the \CRANpkg{JM} package and elsewhere
\citep[\citet{andersen_statistical_1993}]{skrondal_generalized_2004}. It
consists of a randomized trial in which 488 subjects were split into two
treatment arms (prednisone \emph{versus} placebo). Repeated measures of
prothrombin ratio were collected over time as well as time to death. The
longitudinal part of the joint model included a linear trajectory with
time in the study, an indicator of first measurement and their
interaction with treatment group. Were also included correlated
individual random effects on the intercept and the slope with time. The
survival part was a proportional hazard model adjusted for treatment
group as well as the dynamics of the longitudinal outcome either through
the current value of the marker or its slope or both. The baseline risk
function was approximated by B-splines with one internal knot. The total
number of parameters to estimate was 17 or 18 (10 for the longitudinal
submodel, and 7 for the survival submodel considering only the curent
value of the marker or its slope or 8 for the survival model when both
the current level and the slope were considered). The marker initially
ranged from 6 to 176 (mean=79.0, sd=27.3).

To investigate the consistency of the results to different dimensions of the marker, we
also considered cases where the marker was rescaled by a factor 0.1 or
10. In these cases, the log-likelihood was rescaled a posteriori to the
original dimension of the marker to make the comparisons possible. The
starting point was systematically set at the default initial value of
the \code{jointModel} function, which is the estimation point obtained
from the separated linear mixed model and proportional hazard model. 

\rev{In addition to EM and BFGS included in \code{JM} package, we also compared the MLA performances with those of the parallel implementation of the L-BFGS-B algorithm provided by the \CRANpkg{optimParallel} package.} Codes and dataset used in this section are available at
\url{https://github.com/VivianePhilipps/marqLevAlgPaper}.

\rev{MLA and L-BFGS-B ran on 3 cores. MLA} converged when the three criteria defined in
section \ref{sec:criteria} were satisfied with tolerance 0.0001, 0.0001
and 0.0001 for the parameters, the likelihood and the RDM, respectively. BFGS and L-BFGS-B converged when the convergence criterion on the log-likelihood was satisfied with the square root of the tolerance of the machine
(\(\approx 10^{-8}\)). The EM algorithm converged when stability on the
parameters or on the log-likelihood was satisfied with tolerance 0.0001
and around \(10^{-8}\) (i.e., the square root of the tolerance of the
machine), respectively.

\begin{table}[ht]
\centering
\begingroup\footnotesize
\begin{tabular}{rrrrrrrr}
  \toprule  Nature of & Algorithm & Scaling & Rescaled log-& Variation of& Variation of& Number of & Time in\\   dependency & & factor & likelihood & value (\%) & slope (\%) & iterations & seconds \\ \midrule
 value & BFGS & 1 & -13958.55 & -3.73 &  & 120 & 27.39 \\ 
 value & BFGS & 0.1 & -13957.91 & -0.01 &  & 490 & 116.33 \\ 
 value & BFGS & 10 & -13961.54 & -9.28 &  & 91 & 18.16 \\ 
 value & LBFGSB & 1 & -13958.41 & -3.56 &  & 289 & 79.07 \\ 
 value & LBFGSB & 0.1 & -13957.69 & -0.11 &  & 244 & 67.53 \\ 
 value & LBFGSB & 10 & error &  &  &  &  \\ 
 value & EM & 1 & -13957.91 & -0.29 &  & 66 & 72.44 \\ 
 value & EM & 0.1 & -13957.72 & 0.14 &  & 104 & 106.70 \\ 
 value & EM & 10 & -13957.94 & -0.59 &  & 62 & 67.80 \\ 
 value & MLA & 1 & {\bf-13957.69} & -0.00 &  & 7 & 34.37 \\ 
 value & MLA & 0.1 & -13957.69 & -0.00 &  & 6 & 29.48 \\ 
 value & MLA & 10 & -13957.69 & -0.00 &  & 17 & 75.48 \\ 
 slope & BFGS & 1 & -13961.41 &  & -1.85 & 251 & 52.76 \\ 
 slope & BFGS & 0.1 & -13961.23 &  & -1.37 & 391 & 87.61 \\ 
 slope & BFGS & 10 & -13980.90 &  & -13.98 & 444 & 80.16 \\ 
 slope & LBFGSB & 1 & -13960.69 &  & -0.15 & 266 & 60.29 \\ 
 slope & LBFGSB & 0.1 & -13960.70 &  & -0.27 & 206 & 47.87 \\ 
 slope & LBFGSB & 10 & -13962.56 &  & -2.87 & 823 & 182.20 \\ 
 slope & EM & 1 & -13960.69 &  & 0.17 & 170 & 161.64 \\ 
 slope & EM & 0.1 & -13960.69 &  & 0.02 & 208 & 196.68 \\ 
 slope & EM & 10 & -13960.70 &  & 0.08 & 156 & 159.58 \\ 
 slope & MLA & 1 & {\bf-13960.69} &  & -0.00 & 11 & 48.00 \\ 
 slope & MLA & 0.1 & -13960.69 &  & -0.00 & 11 & 48.10 \\ 
 slope & MLA & 10 & -13960.69 &  & 0.00 & 14 & 61.61 \\ 
 both & BFGS & 1 & -13951.60 & 15.97 & -28.17 & 164 & 37.83 \\ 
 both & BFGS & 0.1 & -13949.82 & 2.66 & -4.63 & 502 & 132.84 \\ 
 both & BFGS & 10 & -13965.25 & 40.31 & -95.26 & 52 & 10.48 \\ 
 both & LBFGSB & 1 & -13950.04 & -1.67 & 7.10 & 800 & 177.61 \\ 
 both & LBFGSB & 0.1 & -13949.42 & -0.01 & 0.38 & 411 & 93.31 \\ 
 both & LBFGSB & 10 & -13985.72 & 67.33 & -147.30 & 18 & 7.75 \\ 
 both & EM & 1 & -13949.82 & 4.10 & -7.22 & 159 & 186.69 \\ 
 both & EM & 0.1 & -13949.44 & 1.68 & -3.66 & 156 & 152.89 \\ 
 both & EM & 10 & -13950.46 & 10.67 & -16.31 & 142 & 220.07 \\ 
 both & MLA & 1 & {\bf-13949.42} & -0.00 & -0.00 & 10 & 49.91 \\ 
 both & MLA & 0.1 & -13949.42 & -0.00 & 0.00 & 10 & 51.63 \\ 
 both & MLA & 10 & -13949.42 & -0.00 & 0.00 & 24 & 121.69 \\ 
  \bottomrule
\end{tabular}
\endgroup
\caption{Comparison of the convergence obtained by MLA, BFGS, \rev{LBFGSB} and EM algorithms for the estimation of a joint model for prothrobin repeated marker (scaled by 1, 0.1 or 10) and time to death when considering a dependency on the current level of prothrobin ('value') or the current slope ('slope') or both ('both'). All the models converged correctly according to the algorithm outputs. We report the final log-likelihood rescaled to scaling factor 1 (for comparison), the  percentage of variation of the association parameters ('value' and 'slope' columns) compared to the one obtained with the overall maximum likelihood with scaling 1, the number of iterations and the running time in seconds. } 
\label{tab:prothro}
\end{table}

Table \ref{tab:prothro} compares the convergence obtained by using the
\rev{four} optimization methods, when considering a pseudo-adaptive
Gauss-Hermite quadrature with 15 points. All the algorithms converged
correctly according to the programs except one with L-BFGS-B which gave an error (non-finite value) during optimization. Although the model for a given
association structure is exactly the same, some differences were
observed in the final maximum log-likelihood (computed in the original
scale of prothrombin ratio). The final log-likelihood obtained by MLA
was always the same whatever the outcome's scaling, showing its
consistency. It was also higher than the one obtained using the three
other algorithms, showing that BFGS, L-BFGS-B and, to a lesser extent, EM did not
systematically converge toward the effective maximum. The difference
could go up to 20 points of log-likelihood for BFGS in the example with
the current slope of the marker as the association structure. The
convergence also differed according to outcome's scaling with BFGS/L-BFGS-B and
slightly with EM, even though in general the EM algorithm seemed
relatively stable in this example. The less stringent convergence of
BFGS/L-BFGS-B and, to a lesser extent, of EM had also consequences on the
parameters estimates as roughly illustrated in Table \ref{tab:prothro}
with the percentage of variation in the association parameters of
prothrombin dynamics estimated in the survival model (either the current
value or the current slope) in comparison with the estimate obtained
using MLA which gives the overall maximum likelihood. The better
performances of MLA was not at the expense of the number of iterations
since MLA converged in at most 22 iterations, whereas several hundreds
of iterations could be required for EM or BFGS. Note however that one
iteration of MLA is much more computationally demanding.

Finally, for BFGS, the problem of convergence was even more apparent when
the outcome is scaled by a factor 10. Indeed, the optimal log-likelihood
of the model assuming a bivariate association structure (on the current
level and the current slope) was worse than the optimal log-likelihood of
its nested model which assumes an association structure only on the
current level (i.e., constraining the parameter for the current slope to
0). We faced the same situation with the L-BFGS-B algorithm when comparing the log-likelihoods with a bivariate association and with an association through the current slope only.

\hypertarget{concluding-remarks}{%
\section{Concluding remarks}\label{concluding}}

We proposed in this paper a general-purpose optimization algorithm based
on a robust Marquardt-Levenberg algorithm. The program, written in
R and Fortran90, is available in \CRANpkg{marqLevAlg}
R package. It provides a very nice alternative to other
optimization packages available in R software such as
\CRANpkg{optim}, \CRANpkg{roptim} \citep{pan_2020} or \CRANpkg{optimx}
\citep{nash_2011} \rev{for addressing complex optimization problems}. In particular, \rev{as shown in our examples, notably the}
estimation of joint models, it is more reliable than classical
alternatives (in particular EM, BFGS and L-BFGS-B). This is due to the very good convergence
properties of the Marquardt-Levenberg algorithm associated with very
stringent convergence criteria based on the first and second derivatives
of the objective function which avoids spurious convergence at saddle
points \citep{commenges_rvs_2006}.

The Marquardt-Levenberg algorithm is known for its very computationally
intensive iterations due to the computation of the first and second
derivatives. However, compared to other algorithms, it converges
in a very small number of iterations (usually less than 30 iterations). \rev{This may not make MLA competitive in terms of running time in simple and rapid settings}. However, the parallel computations of the derivatives can largely speed up the program and make it very competitive with
alternatives in terms of running time \rev{in complex settings}. 

We chose in our implementation to rely on RDM criterion which is a very
stringent convergence criterion. As it is based on the inverse of the
Hessian matrix, it may cause non-convergence issues when some parameters
are at the border of the parameter space (for instance 0 for a parameter
constrained to be positive). In that case, we recommend to fix the
parameter at the border of the parameter space and run again the
optimization on the rest of the parameters. In cases where the
stabilities of the log-likelihood and of the parameters are considered
sufficient to ensure satisfactory convergence, the program outputs might
be interpreted despite a lack of convergence according to the RDM, \rev{as is done for other algorithms that only converge according to parameter and/or objective function stability. }

As any other optimization algorithm based on the steepest descent, MLA
\rev{is a local optimizer}. It does not ensure the convergence of multimodal objective functions toward
the global optimum. In such a context we recommend the use of a grid search which consists in running the algorithm from a grid of (random) initial values and retaining the best result as the final solution. We illustrate in supplementary material how this technique succeeds in finding the global minimum with the Wild function of the \code{optim} help page.

\CRANpkg{marqLevAlg} is not the first optimizer to exploit parallel computations. Other R optimizers include a parallel mode, in particular stochastic optimization packages like \CRANpkg{DEoptim} \citep{DEoptim_2011}, \CRANpkg{GA} \citep{ga_2017}, \CRANpkg{rgenoud} \citep{rgenoud_2011} or  \CRANpkg{hydroPSO} \citep{hydroPSO_2020}. We compared these packages, the local optimizer of \code{optimParallel}, and \CRANpkg{marqLevAlg} for the estimation of the linear mixed model described in Section \ref{example}. For this specific problem \code{marqLevAlg} was the fastest, followed by \code{optimParallel} (results shown in supplementary files).

With its
parallel implementation of derivative calculations combined with very
good convergence properties of MLA, \CRANpkg{marqLevAlg} package provides a
promising solution for the estimation of complex statistical models in
R. We have chosen for the moment to parallelize the
derivatives which is very useful for optimization problems involving
many parameters. However we could also easily parallelize the
computation of the objective function when the latter is decomposed into
independent sub-computations as is the log-likelihood computed
independently on the statistical units. This alternative is currently
under development.

\hypertarget{funding}{%
\section*{Funding}\label{funding}}

This work was funded by French National Research Agency {[}grant number
ANR-18-CE36-0004-01 for project DyMES{]} and {[}grant number
ANR-2010-PRPS-006 for project MOBYDIQ{]}.

\hypertarget{acknowlegdments}{%
\section*{Acknowlegdments}\label{acknowlegdments}}

The computing facilities MCIA (Mésocentre de Calcul Intensif Aquitain)
at the Université de Bordeaux and the Université de Pau et des Pays de
l'Adour provided advice on parallel computing technologies, as well as
computer time.

\bibliography{philipps.bib}


\address{
    Viviane Philipps\\
  Inserm, Bordeaux Population Health Research Center, UMR 1219,\\
  Univ. Bordeaux, F-33000 Bordeaux, France\\
  146 rue Léo Saignat\\
  33076 Bordeaux Cedex\\
  France\\
  E-mail: \email{viviane.philipps@u-bordeaux.fr}
}

\address{
  Boris P. Hejblum\\
  Inserm, Bordeaux Population Health Research Center, UMR 1219,\\
  Inria BSO SISTM, \\
  Vaccine Reserch Institute (VRI), F-94000 Créteil, France \\
  Univ. Bordeaux, F-33000 Bordeaux, France\\
  146 rue Léo Saignat\\
  33076 Bordeaux Cedex\\
  France\\
  E-mail: \email{boris.hejblum@u-bordeaux.fr}
}

\address{
  Mélanie Prague\\
  Inserm, Bordeaux Population Health Research Center, UMR 1219,\\
  Inria BSO SISTM, \\
  Vaccine Reserch Institute (VRI), F-94000 Créteil, France \\
  Univ. Bordeaux, F-33000 Bordeaux, France\\
  146 rue Léo Saignat\\
  33076 Bordeaux Cedex\\
  France\\
  E-mail: \email{melanie.prague@inria.fr}
}

\address{
  Daniel Commenges\\
  Inserm, Bordeaux Population Health Research Center, UMR 1219,\\
  Inria BSO SISTM, \\
  Univ. Bordeaux, F-33000 Bordeaux, France\\
  146 rue Léo Saignat\\
  33076 Bordeaux Cedex\\
  France\\
  E-mail: \email{daniel.commenges@u-bordeaux.fr}
}

\address{
  Cécile Proust-Lima\\
  Inserm, Bordeaux Population Health Research Center, UMR 1219,\\
  Univ. Bordeaux, F-33000 Bordeaux, France\\
  146 rue Léo Saignat\\
  33076 Bordeaux Cedex\\
  France\\
  E-mail: \email{cecile.proust-lima@inserm.fr}
}

\newpage

\hypertarget{appendix}{%
\section*{Appendix}\label{appendix}}
\addcontentsline{toc}{section}{Appendix}

\hypertarget{a1.-standard-examples-from-the-litterature}{%
\subsection*{A1. Standard examples from the
literature}\label{a1.-standard-examples-from-the-litterature}}
\addcontentsline{toc}{subsection}{A1. Standard examples from the
litterature}

We assessed the performances of our implementation of the
Marquardt-Levenberg algorithm by following the strategy of
\citet{more_1981} for testing algorithms in unconstrained problems. They
provide a series of 35 objective functions along with initial values. We
used for this purpose the R package funconstrain. The 35 problems were
optimized with marqLevAlg and with 6 other usual methods: Nelder-Mead,
BFGS, CG, L-BFGS-B (all 4 from the optim function), L-BFGS-B from
optimParallel package, and nlminb.

\begin{figure}
{\centering \includegraphics[width=1\linewidth]{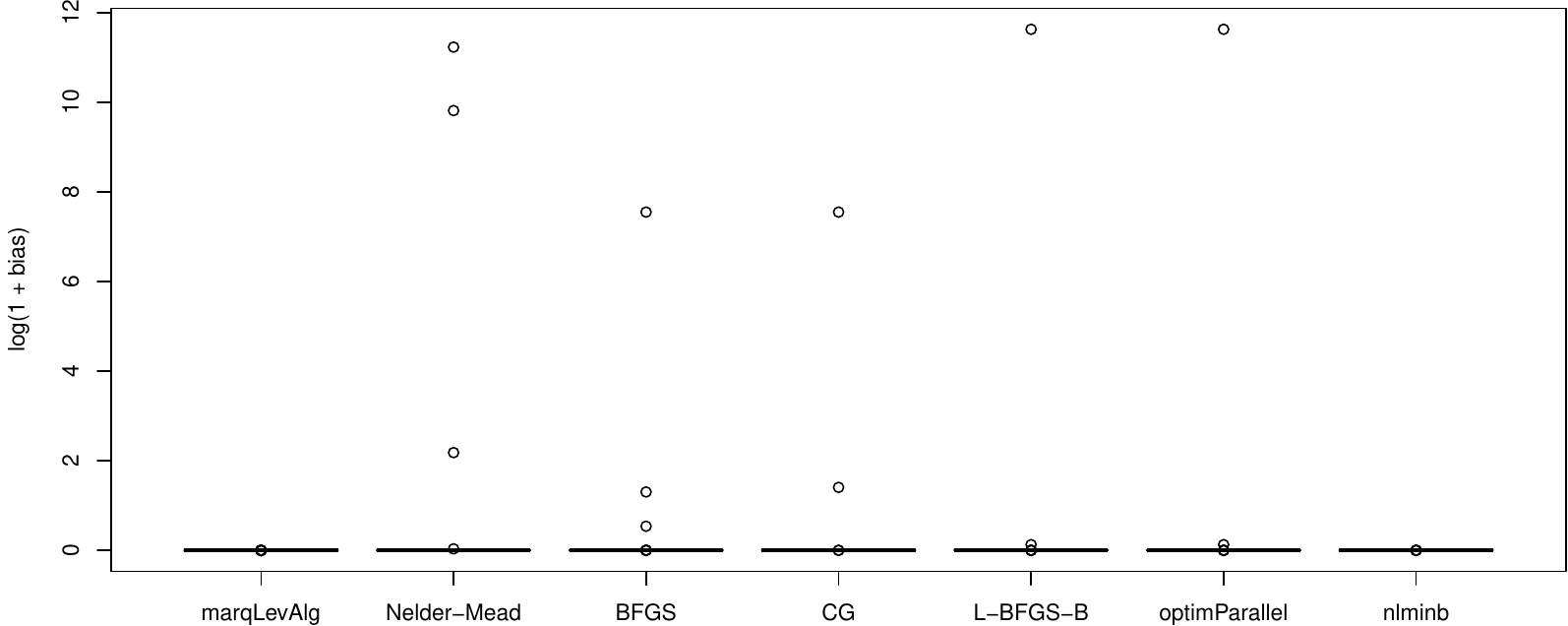} 
}
\caption[Log-scaled bias between real solution and final optimum value at convergence point for 7 different algorithms ]{Log-scaled bias between real solution and final optimum value at convergence point for 7 different algorithms : marqLevAlg, Nelder-Mead, BFGS, CG, L-BFGS-B, optimParallel and nlminb}\label{fig:figure35ex}
\end{figure}

Table \ref{tab:table35ex} shows absolute differences between the real
minimum of the objective function and the result obtain by each
algorithm. Blanks indicate no convergence of the algorithm or error. The
differences are also plotted in the log scale in figure
\ref{fig:figure35ex}.

\begin{table}[h]
\centering
\begin{tabular}{lrrrrrrr}
  \hline
 & marqLevAlg & Nelder-Mead & BFGS & CG & L-BFGS-B & optimParallel & nlminb \\ 
  \hline
rosen & 0.000 & 0.000 & 0.000 &  & 0.000 & 0.000 & 0.000 \\ 
  freud\_roth & 0.000 & 0.000 & 0.000 &  & 0.000 & 0.000 & 0.000 \\ 
  powell\_bs & 0.000 & 0.000 &  &  & 0.135 & 0.135 &  \\ 
  brown\_bs &  & 18271 & 0.000 &  & 0.000 & 0.000 & 0.000 \\ 
  beale & 0.000 & 0.000 & 0.000 & 0.000 & 0.000 & 0.000 & 0.000 \\ 
  jenn\_samp & 0.000 & 0.000 & 1896 & 1896 &  &  & 0.000 \\ 
  helical & 0.000 & 0.000 & 0.000 & 0.000 & 0.000 & 0.000 & 0.000 \\ 
  bard & 0.000 & 0.000 & 0.000 &  & 0.000 & 0.000 & 0.000 \\ 
  gauss & 0.000 & 0.000 & 0.000 & 0.000 & 0.000 & 0.000 & 0.000 \\ 
  meyer &  & 75306 & 0.002 &  & 112035 & 112035 &  \\ 
  gulf & 0.000 & 0.000 &  &  & 0.000 & 0.000 & 0.000 \\ 
  box\_3d & 0.000 & 0.000 & 0.000 &  & 0.000 & 0.000 & 0.000 \\ 
  powell\_s & 0.000 & 0.000 & 0.000 &  & 0.000 & 0.000 & 0.000 \\ 
  wood & 0.000 & 7.855 & 0.000 &  & 0.000 & 0.000 & 0.000 \\ 
  kow\_osb & 0.000 & 0.000 & 0.000 &  & 0.000 & 0.000 & 0.000 \\ 
  brown\_den & 0.0 & 0.0 & 0.0 & 0.0 & 0.0 & 0.0 & 0.0 \\ 
  osborne\_1 & 0.000 & 0.000 & 0.000 &  & 0.000 & 0.000 & 0.000 \\ 
  biggs\_exp6 & -0.006 &  &  &  & -0.000 & -0.000 & -0.000 \\ 
  osborne\_2 & 0.000 &  & 0.000 &  & 0.000 & 0.000 & 0.000 \\ 
  watson & 0.000 &  & 0.000 &  & 0.000 & 0.000 & 0.000 \\ 
  ex\_rosen & 0.000 &  & 0.000 & 0.000 & 0.000 & 0.000 & 0.000 \\ 
  ex\_powell & 0.000 &  & 0.000 & 0.000 & 0.000 & 0.000 &  \\ 
  penalty\_1 & 0.000 & 0.000 & 0.000 &  & 0.000 & 0.000 & 0.000 \\ 
  penalty\_2 & 0.000 & 0.000 & 0.000 & 0.000 & 0.000 & 0.000 &  \\ 
  var\_dim & 0.000 &  & 0.000 & 0.000 & 0.000 & 0.000 & 0.000 \\ 
  trigon & 0.000 &  & 0.000 & 0.000 & 0.000 & 0.000 & 0.000 \\ 
  brown\_al & 0.000 &  & 0.000 & 0.000 & 0.000 & 0.000 & 0.000 \\ 
  disc\_bv & 0.000 &  & 0.000 &  & 0.000 & 0.000 & 0.000 \\ 
  disc\_ie & 0.000 &  & 0.000 & 0.000 & 0.000 & 0.000 & 0.000 \\ 
  broyden\_tri & 0.000 &  & 0.713 & 0.000 & 0.000 & 0.000 & 0.000 \\ 
  broyden\_band & 0.000 &  & 2.680 & 3.076 & 0.000 & 0.000 & 0.000 \\ 
  linfun\_fr & 0.000 &  & 0.000 & 0.000 & 0.000 & 0.000 & 0.000 \\ 
  linfun\_r1 &  &  & 0.000 & 0.000 & 0.000 & 0.000 & 0.000 \\ 
  linfun\_r1z &  &  & 0.000 & 0.000 & 0.000 & 0.000 & 0.000 \\ 
  chebyquad & 0.000 &  & 0.000 & 0.000 & 0.000 & 0.000 & 0.000 \\ 
   \hline
\end{tabular}
\caption{Absolute bias, for each of the 35 problems, between the real solution and the final optimum value at convergence point for 7 different algorithms : marqLevAlg, Nelder-Mead, BFGS, CG, L-BFGS-B, optimParallel and nlminb. An empty case means that the algorithm did not converged.} 
\label{tab:table35ex}
\end{table}

MarqLevAlg converged in 31 of the 35 cases and found the optimum of the objective
function with minimal bias. Except for nlminb which showed the same very
good performances (31 convergences out of the 35 cases and no bias), the other algorithms converged at least once very far
from the effective objective value. In addition, Nelder-Mead and CG
algorithms converged only in approximately half of the cases. This
illustrates the reliability of marLevAlg to find the optimum in
different settings.

\hypertarget{a2.-marquardt-levenberg-implementations-for-nonlinear-least-square-problems}{%
\subsection*{A2. Marquardt-Levenberg implementations for nonlinear least
square
problems}\label{a2.-marquardt-levenberg-implementations-for-nonlinear-least-square-problems}}
\addcontentsline{toc}{subsection}{A2. Marquardt-Levenberg
implementations for nonlinear least square problems}

Although not restricted to nonlinear least square problems, we compared
our implementation of Marquardt-Levenberg algorithm with two other
implementations dedicated to nonlinear least square problems in the R
packages nlmrt and minpack.lm. We used the examples given in those two
packages to compare our results to the one obtained by the two other
implementations. We compared the implementations in terms of residual
sum-of-squares (RSS) at convergence and runtime in microseconds (as the
mean runtime over 100 replicates).

Table \ref{tab:exnlmrt} summarizes the results of the three examples
provided by the help page of nlmrt. The Hobbs problem has been run
several times with different initial values, in a scaled or an unscaled
version, and with an analytical gradient. The examples were tested with
function nlxb (or nlfb when the analytical gradient was specified) and
function marqLevAlg (in sequential mode). Table \ref{tab:exminpack}
summarizes the results obtained on the two examples provided in the help
of the nls.lm function from minpack.lm package.

These tests show that our implementation provides the same final RSS as
the two other implementations in these examples. We note that one run
did not converge with marqLevAlg. Our implementation was yet
systematically longer than the two others. We did expect this as our
implementation is not dedicated to such simple situations but rather to
complex optimization problems as shown in other examples in the main
manucript (e.g., linear mixed model, joint model, latent class model).

\begin{table}[ht]
\centering
\begin{tabular}{lrrrr}
  \hline   & \multicolumn{2}{c}{nlxb/nlfb} & \multicolumn{2}{c}{marqLevAlg}\\ \hline
 & objective function & runtime & objective function & runtime \\ 
  \hline
One parameter problem & 8.9296 & 1731 & 8.9296 & 775 \\ 
  Hobbs problem unscaled - start1 & 2.5873 & 13047 &  & 155910 \\ 
  Hobbs problem unscaled - easy & 2.5873 & 3522 & 2.5873 & 34805 \\ 
  Hobbs problem scaled - start1 & 2.5873 & 7763 & 2.5873 & 12020 \\ 
  Hobbs problem scaled - easy & 2.5873 & 3615 & 2.5873 & 9358 \\ 
  Hobbs problem scaled - hard & 2.5873 & 16344 & 2.5873 & 27588 \\ 
  Hobbs problem scaled - start1 - gradient & 2.5873 & 3422 & 2.5873 & 13455 \\ 
  Gabor Grothendieck problem & 0.0000 & 1871 & 0.0000 & 983 \\ 
   \hline
\end{tabular}
\caption{Final objective function value and runtimes (in microseconds) of least squares problems solved with nlmrt and marqLevAlg packages.} 
\label{tab:exnlmrt}
\end{table}

\begin{table}[ht]
\centering
\begin{tabular}{lrrrr}
  \hline   & \multicolumn{2}{c}{nls.lm} & \multicolumn{2}{c}{marqLevAlg}\\ \hline
 & objective function & runtime & objective function & runtime \\ 
  \hline
Example1 & 0.7986 & 313 & 0.7986 & 7062 \\ 
  Example2 & 79237 & 2374 & 79237 & 38584 \\ 
  Example2 - gradient & 79237 & 2388 & 79237 & 20307 \\ 
   \hline
\end{tabular}
\caption{Final objective function value and runtimes (in microseconds) of least squares problems solved with minpack.lm and marqLevAlg packages.} 
\label{tab:exminpack}
\end{table}

\hypertarget{a3.-other-parallelized-optimization-algorithms}{%
\subsection*{A3. Other parallelized optimization
algorithms}\label{a3.-other-parallelized-optimization-algorithms}}
\addcontentsline{toc}{subsection}{A3. Other parallelized optimization
algorithms}

Other optimizers are available in R with a parallel mode such as
DEoptim, GA, rgenoud, hydroPSO and optimParallel \citep{Mullen_2014}.
Although these algorithms are dedicated to global optimization, we used
them in a local optimization problem to contrast the performances of
marqLevAlg with them. We used the example of estimation of a linear
mixed model presented in the Example section. We estimated the model
with packages rgenoud, DEoptim, hydroPSO, GA and optimParallel using one
and two cores. Runtimes are summarized in Table \ref{tab:parallel}. In
this situation, our algorithm showed by far the minimum runtimes even
though its speed up was slightly less (1.51) than others
(\textgreater{}1.76).

\begin{table}[h]
\begin{tabular}{lrrr}
  \hline
  R function & sequential runtime & parallel runtime & speed  up \\
  \hline
  marqLevAlg    &  21.89&  14.47&1.51 \\
  genoud        & 650.11& 348.18&1.87 \\
  DEoptim       & 318.50& 139.40&2.28 \\
  hydroPSO      & 860.97& 393.07&2.19 \\
  ga            &  94.49&  51.17&1.85 \\
  optimParallel &  63.49&  36.02&1.76 \\
  \hline
\end{tabular}
\caption{Mean runtimes over 10 replicates of the estimation of a linear mixed model using marqLevAlg, rgenoud, DEoptim, hydroPSO, GA and optimParallel packages in sequential mode and in parallel mode using two cores.}
\label{tab:parallel}
\end{table}

\hypertarget{a4.-sensitivity-to-initial-values}{%
\subsection*{A4. Sensitivity to initial
values}\label{a4.-sensitivity-to-initial-values}}
\addcontentsline{toc}{subsection}{A4. Sensitivity to initial values}

\hypertarget{comparison-with-another-marquardt-levenberg-implementation}{%
\subsubsection{Comparison with another Marquardt-Levenberg
implementation}\label{comparison-with-another-marquardt-levenberg-implementation}}

We considered an example from the non-linear least squares area to
compare convergence rates, objective function's final value, and
sensitivity to initial values obtained by marqLevAlg in comparison with
the Marquardt-Levenberg algorithm implementation of minpack.lm package
with nls.lm function.

We estimated the 3-parameter model \(y = a * \exp(x * b) + c\) using 100
starting values drawn uniformly between -10 and 10. The procedure was
replicated on 100 datasets.

Over the 10000 estimations, marqLevAlg converged in 51.55\% of the
cases, whereas 65.98\% of the nls.lm models converged, as shown in table
\ref{tab:VI}. For nls.lm, this mixes the three convergence criteria,
namely according to the objective function stability (value info=1 in
the code), to the parameters stability (info=2) or to both (info=3). A
fourth convergence criterion based on the angle between the objective
function and its gradient was avalaible (info=4) but was never used in
the 10000 runs.

\begin{figure}
{\centering \includegraphics[width=1\linewidth]{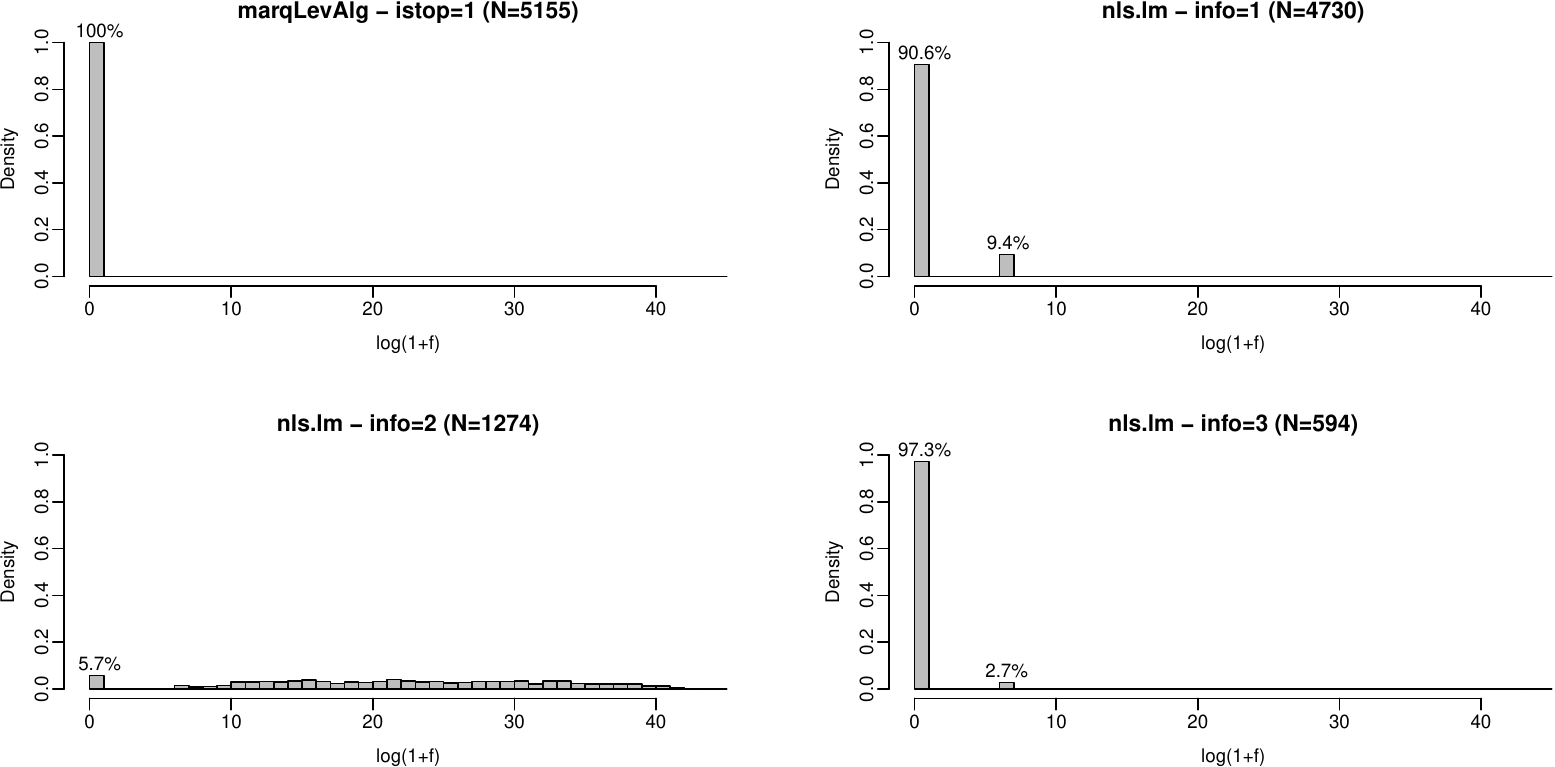} 
}
\caption[Final value of the objective function at convergence for marqLevAlg algorithm and for nls.lm algorithm according to the type of convergence criterion met (1 for objective function, 2 for parameters, 3 for both)]{Final value of the objective function at convergence for marqLevAlg algorithm and for nls.lm algorithm according to the type of convergence criterion met (1 for objective function, 2 for parameters, 3 for both). Are only reported the runs that converged, and results are in the log scale so that small differences are not blurred by some extreme differences. }\label{fig:figureVI}
\end{figure}

While the minimum value was effectively reached for all the convergences
of marqLevAlg, 1660 estimations that converged according to nls.lm were
far from the effective optimum. This reduced the proportion of
satisfying convergences to 49.38\% (so similar rate as marqlevAlg) but
more importantly illustrated the convergence to saddle points when using
classical convergence criteria. These convergences to saddle points are
illustrated in Figure \ref{fig:figureVI}. The problem of spurious
convergence was observed in all the types of convergence although it was
particularly important when nls.lm converged with the paramater
stability criterion (an extreme value was obtained in 443 and 16 runs
for convergence on the parameters and on the function, respectively).

\begin{table}[h]
\begin{tabular}{ll|r|r|r}
 & & \multicolumn{2}{c|}{marqLevAlg} \\
  && convergence & non convergence & total \\
  \hline
 & convergence in function   &4262&  468  & \\
nls.lm&convergence in parameters &  87& 1187 & 6598\\
&convergence in both       & 591&    3 & \\
\cline{2-5}
&  non convergence           & 215& 3187 & 3402\\
  \hline
&  total & 5155 & 4845 & 10000
\end{tabular}
\caption{Summary of the convergence status of marqLevAlg and nls.lm over 10000 runs (100 simulated samples and 100 different starting points each).}
\label{tab:VI}
\end{table}

\hypertarget{global-optimization-using-grid-search}{%
\subsubsection{Global optimization using grid
search}\label{global-optimization-using-grid-search}}

The Marquardt-Levenbergh algorithm performs local optimization. In
situations were a global minimum (or maximum) is sought, the algorithm
can still be used with a grid search. It consists in running the
algorithm with multiple different initial values and retaining the best
result.

\begin{figure}
{\centering \includegraphics[width=1\linewidth]{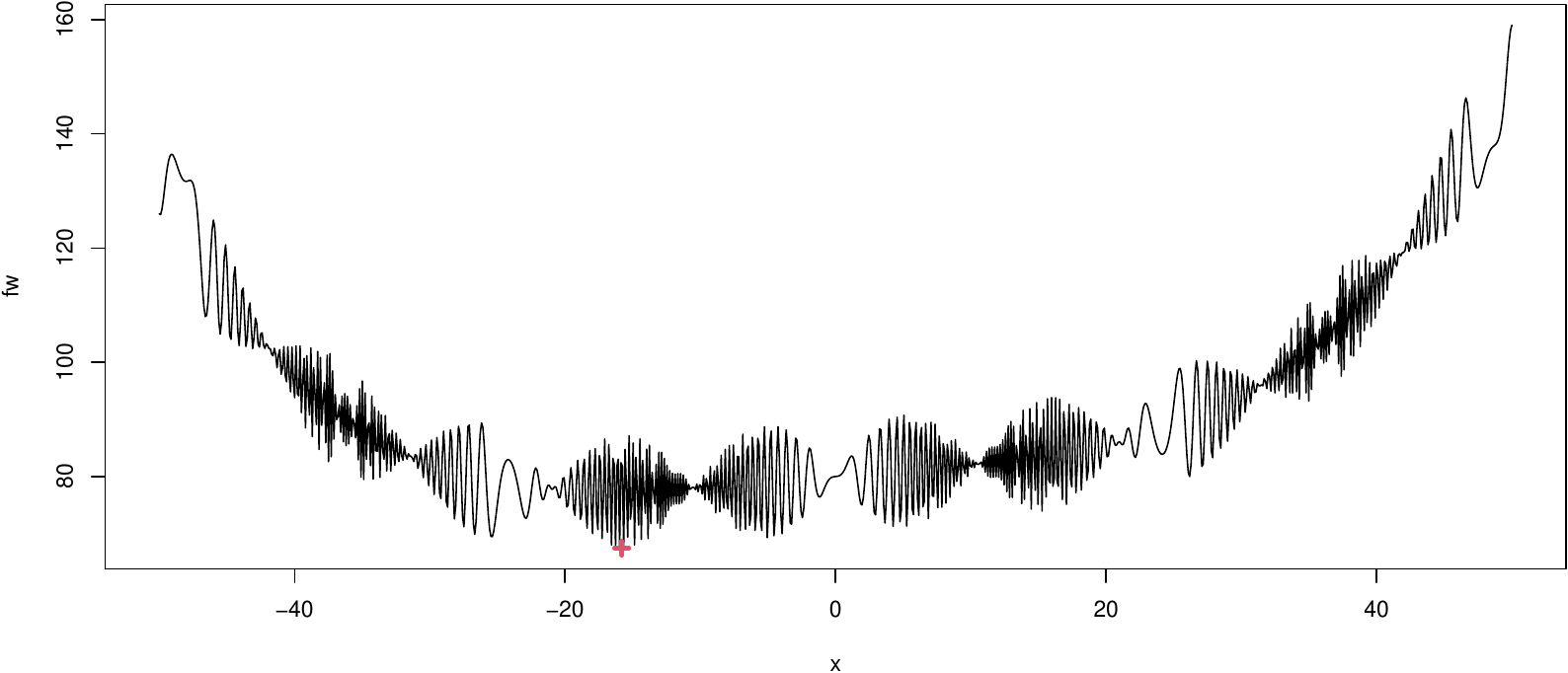} 
}
\caption[The Wild function of the help page of optim function]{The Wild function of the help page of optim function. Global minimum appears in red.}\label{fig:figureWild}
\end{figure}

We illustrate this with the Wild function plotted in Figure
\ref{fig:figureWild} and defined as:
\[fw(x) = 10 * \sin(0.3 * x) * \sin(1.3 * x^2) + 0.00001 * x^4 + 0.2 * x + 80\]
This function is given as an example in the help page of the optim
function for global optimization problem.

We ran the marqLevAlg algorithm 200 times from starting points defined
by a regular grid between values -50 and 50. The minimum value over the
200 trials did coincide with the results of the global optimization
algorithm SANN as shown in table \ref{tab:resgridsearch}.

\begin{table}[h!]
\centering
\begin{tabular}{lrr}
  \hline
 & SANN & grid search MLA \\ 
  \hline
minimum & 67.4680 & 67.4677 \\ 
  param & -15.8153 & -15.8152 \\ 
   \hline
\end{tabular}
\caption{Optimization results on the Wild function with algorithm SANN and MLA} 
\label{tab:resgridsearch}
\end{table}

\end{article}

\end{document}